\documentclass[12pt,article]{amsart2000}
\usepackage{amsfonts}
\RequirePackage{amsmath2000}
\newcommand{\crr} {{\mathcal{R}}}            \newcommand{\cc} {{\mathcal{C}}}
\newcommand{\ca} {{\mathcal{A}}}             
\newcommand{\cl} {{\mathcal{L}}}             
\newcommand{\ce} {{\mathcal{E}}}             \newcommand{\cg} {{\mathcal{G}}}
\newcommand{\cu} {{\mathcal{U}}}             \newcommand{\al} {{\alpha}}
\newcommand{\be} {{\beta}}              
\newcommand{\Ga} {{\mathit\Gamma}}         \newcommand{\de} {{\delta}}

             \newcommand{\om} {{\omega}}
\newcommand{\Om} {{\varOmega}}          
        
        \newcommand{\bn} {{\mathbb N}}
          \newcommand{\ba} {{\mathbb A}}
\newcommand{\bc} {{\mathbb C}}          \newcommand{\br} {{\mathbb R}}

\begin{document}

\title{\bf Remarks on ``singularities''}

\author{\bf Anastasios Mallios}

\date{}

\maketitle

\pagestyle{myheadings} \markboth{\centerline {\small {\sc
{Anastasios Mallios}}}}
         {\centerline {\small {\sc {Remarks on ``singularities''}}}}

%-----------------------------------------------------------------
\begin{quote}{\footnotesize {\bf Abstract.} We present herewith certain thoughts
on the important subject of nowadays physics, pertaining to the
so-called {\it ``singularities''}, that emanated from looking at
the theme in terms of ADG (:\;{\it abstract differential
geometry}). Thus, according to the latter perspective, we can
involve ``singularities'' in our arguments, while still employing
fundamental differential-geometric notions such as connections,
curvature, metric and the like, retaining also the form of
standard important relations of the classical theory (e.g.
Einstein and/or Yang-Mills equations, in vacuum), even within that
generalized context of ADG. To wind up, we can {\it extend} (in
point of fact, {\it calculate}) {\it over singularities classical
differential-geometric relations/equations, without altering their
forms and/or changing the standard arguments}\,; the change
concerns thus only {\it the way}, we employ the usual differential
geometry of smooth manifolds, so that the base {\it ``space''}
acquires now quite a {\it secondary r\^{o}le, not contributing} at
all (!) {\it to the differential-geometric technique/mechanism}
that we apply. Thus, the latter by definition refers directly to
the {\it objects} being involved---the objects that ``live on that
space'', which {\it by themselves are not}, of course,
i\;p\;s\;o\;\;f\;a\;c\;t\;o {\it ``singular''}\,!}
\end{quote}

{\bf 0.} According to the {\it Principle of General Relativity},
physical objects are, so to say, {\it ``differentiable objects''},
in the sense that {\it General Relativity} is rooted, roughly
speaking, as a mathematical-physical theory, on the classical
differential geometry of {\it differential} (so-called {\it
``smooth''}) {\it manifolds}. Thus, in view of the above
principle, (see, for instance, M.~Nakahara [28: p. 28]),

\bigskip
\noindent (0.1) \hfill
\begin{minipage} {11cm}
{\it ``all laws in physics take the same form in any coordinate
system''.}
\end{minipage}

\bigskip
\noindent Yet, in accord with (0.1), it is also the so-called {\it
``gauge principle''} (ibid., p. 10), viz. we still adopt that;

\bigskip
\noindent (0.2)\hfill \begin{minipage}{11cm} {\it ``physics should
not depend on how we describe it''.}
\end{minipage}

\bigskip \noindent
see also e.g. R.\;Torretti [37: p. 65]. Thus, in other words, the
laws of Nature are\linebreak independent of any particular
coordinate system (: yet, {\it ``laboratory''}, or even {\it
``local gauge''}), by means of which we virtually effectuate, each
time (yet, make computations about) these laws\,!

Now, {\it classical differential geometry} (CDG) cannot be
applied, by its very de\-fi\-ni\-tion, on non-smooth objects, so
that, if we are still going to employ methods of CDG to {\it
``non-smooth situations''}, we are thus compelled to concoct other
means to overcome such type of impediments of the theory
concerned. So it was exactly here that the mathematicians started
to feel the need of developing techniques, similar to those of CDG
(it seems they were already convinced of the effectiveness of the
latter theory, as it actually happened, of course (!)), which,
however, now should be able to cope with the so-called {\it
``singularities''} phenomena.

In this context, we see thus {\it two main tendencies}, during the
last few decades: The {\it first} (older) {\it one} imitates
methods of CDG, that now could be applied already on not
necessarily smooth functions, sticking, however, more or less, at
notions of the classical theory, as, for instance, tangent
vectors, hence, tangent spaces, vector fields, therefore,
differential forms, as well, and the like. We can call this point
of view {\it Polish-German school} (thus, R.~Sikorski, M.~Heller,
W.~Sasin et al., and K.~Spallek, Buchner et al.). {\it The other}
(more recent) {\it one}, presumably influenced, by {\it quantum
theory} problems, is mainly concentrated on {\it
``non-commutative''} notions, calculations etc, hence, forced also
to develop a {\it ``non-commutative differential geometry''} ({\it
French school}, A.~Connes et al.). In this concern, we further
remark that, both of the previous perspectives try to apply
extended (generalized) methods of CDG to non-classical
(differential-geometric) frameworks, as, for example, in a {\it
``general-relativistic set-up with singularities''}.

Within the same context, we should mention here what we may call
the {\it American school}, appeared about the same period, with
the first one, as above, started by N.~Aronszjan and further
continued by C.D.~Marshall, J.W.~Smith, K.T.~Chen, M.A.~Mostow et
al. In this regard, we should also notice an early appearance
(1956) of such a perspective, as before, {\it of generalizing the
notion of a differential manifold}, by the {\it Japanese}
mathematician I.~Satake (therefore, the so-called {\it ``Satake
manifolds''}, or even {\it ``$V$-manifolds''}).

On the other hand, we have the recently developed {\it ``abstract
differential geometry''} (ADG). Here, quite independently of all
the previous methods, this technique started, at a first stage,
from {\it topological algebra theory} notions, entangled already
in an intense geometrical context (we also remark here that the
totality of scalar-valued differentiable functions on a given
smooth manifold is, in effect, an important {\it (non-normed !)
topological algebra}), and finally concluded to entirely algebraic
(no more topological-algebraic) concepts; the crucial instrument
herewith was {\it sheaf theo\-ry} (of course, the methods of the
latter theory are also algebraic), as well as, {\it sheaf
co\-ho\-mo\-lo\-gy}. In point of fact, one aims here at developing
classical differential geometry, i\;n\;\;a\;b\;s\;r\;a\;c\;t\;o,
where {\it no Calculus}, hence, (local) smoothness of the standard
theory {\it is needed at all}\;! So a general idea herewith is
that;

\bigskip
\noindent (0.3) \hfill
\begin{minipage} {11cm}
whenever we try to abstract a notion of CDG, we have first to find
the pertinent {\it function}, that may represent(:\;be connected
with) it, and then translate it into the appropriate {\it sheaf
morphism}.
\end{minipage}

\bigskip
\noindent The big surprise and vindication, as well, of the
methods of ADG happened quite recently, when realizing that one
could use, as {\it ``domain of numbers''}, alias {\it ``(extended)
arithmetics''} (yet, {\it ``sheaf of coefficients''}) of ADG the
strange (!), however, quite efficient, indeed, algebra of {\it
``generalized functions''} and even, more generally, {\it
``multi-foam algebra''} of such functions, initiated by
E.E.~Rosinger, in effect, the respective {\it sheaves} of these
algebras (see e.g. E.E.\;Rosinger [33], or even
A.\;Mallios-E.E.\;Rosinger [25]. The important thing here is that
the previous algebras of (generalized) functions contain, by their
very definition, a tremendous, in point of fact, by simply
referring to multi-foam algebras, as above, the {\it biggest},
thus far, {\it amount of singularities} (of any ``type''), that
one can consider. In this context, it is still to be noticed that
the aforesaid algebras have already a successful career in
problems, pertaining to {\it non-linear} PDE's. So, in other
words,

\bigskip
\noindent (0.4) \hfill
\begin{minipage} {11cm}
by applying the technique of ADG, when using {\it Rosinger's
algebra sheaf}, as our {\it ``arithmetics''}, we actually derive,
by looking at the corresponding equations (of the theory), {\it
solutions}, which are {\it free of singularities}. Namely, those
singularities that are already contained in our (Rosinger's) {\it
algebra (sheaf) of coefficients}.
\end{minipage}

\bigskip
\noindent One can remark here that something like this was already
a demand of A.~Einstein. That is, we do have now, in that context,
such a

\bigskip
\noindent (0.5) \hfill
\begin{minipage} {11cm}
    {\it ``method ... to derive ... solutions ... free of
    singularities ...'',}
\end{minipage}

\bigskip
\noindent as he actually was looking for (see [6: p. 165]).
Furthermore, another moral of the preceding is that,

\bigskip
\noindent (0.6) \hfill
\begin{minipage} {11cm}
whenever we meet a {\it ``singularity''}, in the classical sense
of the word, one has just to {\it find} {\it the appropriate
algebra of coefficients}, that contains it, provided, of course,
the algebra at issue fits in also well, within the framework of
ADG (take, for instance, the pertinent Rosinger's algebra
(sheaf)).
\end{minipage}

\bigskip
\noindent As an anticipation to our last proposition above, one
may certainly look at the argument of D.~Finkelstein, referring to
the {\it ``past-future asymmetry of the gravitational field''}
(:\;{\it ``Eddington-Finkelstein coordinates''}, see e.g. [27: p.
828]; my thanks are due here to I.\;Raptis, who brought to my
attention the relevant work of Finkelstein). Within the same
framework, concerning the potential significance of {\it
incorporating singularities into the standard equations} (of
general relativity), see, for instances, M.\;Heller [12: p. 924],
along with the relevant Refs therein.

Now, the aforementioned {\it algebras} are, by definition, {\it
commutative}, while, of course, it is actually through them that
we always make our particular {\it calculations}. In this concern,
we should also notice that {\it calculations}, according to a
famous apostrophe of N.\;Bohr, {\it even when referred to a
quantum-theoretic framework, have to be commutative} (!). Yet, by
still paraphrasing the same motto, we can further say that {\it
``our measuring apparatus is a classical object, giving classical
results}, hence, {\it commutative} ones (in point of fact, {\it
eigenvalues}, viz. $c$-{\it numbers}, of the {\it ``observable
operator''}; see also, for instance, R.\;Gilmore [11: p. 71]).

Yet, as a further corroboration of the {\it methods of abstract
differential geometry}, that might also be construed, as being in
accord with the aforementioned {\it ``gauge principle''}
(cf.\;(0.2)), one might refer to the fact that the {\it Einstein
equation} (i\;n\;\;v\;a\;c\;u\;o) has exactly the {\it same form
in the classical theory and in} ADG. (See A\;Mallios [19: p. 89,
(3.11), or even the same author [21: Vol. II; Chapt. IX,
Section\;3, rel. (3.11)]).

\bigskip
{\bf 1.} By commenting upon (0.3), as above, we can still remark
that Leibniz, already at his time, was looking for a {\it
``geometric calculus''}; that is, for a device, {\it ``acting
directly on the geometric objects, without} [at all] {\it the
intervening of coordinates''} (commutative or not!). We recall
here that the same great scientist wrote once to de L'H\^{o}pital
that {\it ``the secret of Analysis lies} [exactly] {\it in an apt
combination of symbols''} (!).

In this context, we can even say that the power of {\it
differential geometry}, the latter being, in effect, an {\it
application of ``differential analysis''} (Calculus$)$ in studying
the {\it ``geometry''} (viz. the {\it inner structure}) of some
{\it ``space''}, is finally proven to be the result of an {\it
inherent mechanism}, supplied by the said device (discipline),
being, in point of fact, {\it independent of any notion of
``space''}. Therefore, a {\it mechanism (:\;``calculus'') referred
to the objects themselves}, that fill up the ``space''. This, of
course, still justifies our endeavor of today to employ that same
mechanism, even in the quantum world (see below {\it ``gauge
theories''}), notwithstanding at that deep there is no, in effect,
any notion of ``space'', in the usual sense of the term (see also,
for instance, C.J.\;Isham [15: p. 400]). However, to be more
precise, the {\it ``space''} is also here. Namely, even at the
quantum deep, as well, {\it the same, as anywhere else\,}; that
is,

\bigskip
\noindent (1.1) \hfill
\begin{minipage} {11cm}
the {\it totality of the} (``geometrical'') {\it objects} (for the
case at hand, the {\it elementary particles}) {\it themselves}
(that fill {\it ``it''} up).
\end{minipage}

\bigskip
\noindent Accordingly, these {\it same objects, viewed}, by virtue
of the preceding, {\it as ``geometrical''} ones, can further be
{\it treated, as such}, too, given that, to this end, {\it we do
not} actually {\it need any space to refer to}, apart from the
objects themselves, that we are observing (detecting). The above
is still another {\it crucial outcome of} ADG. So, practically
speaking, we can say that,

\bigskip
\noindent (1.2) \hfill
\begin{minipage} {11cm}
{\it to perform} ADG, {\it one does not} actually {\it need any
``space'' at all\,!}
\end{minipage}

\bigskip
\noindent Accordingly, one can also look at the above issue, as a
post-anticipation or, at least, as a response to the
aforementioned demand of Leibniz.

Here it is also worthwhile to recall an utterance of B.\;Riemann
in his famous Habilitationsschrift (:\;{\it ``\"{U}ber die
Hypothesen, welche der Geometrie zu Grunde liegen''}, 1854), in
that

\bigskip
\noindent (1.3) \hfill
\begin{minipage} {11cm}
    ``Ma\ss bestimungen erfordern eine Unabh\"{a}ngichkeit der
    Gr\"{o}\ss en vom Ort, die in mehr als einer Weise stattfinden
    kann.'': Spe\-ci\-fi\-ca\-tions of mass [:\;measurements] require an independence
    of quantity from position, which can happen in more than one way.
\end{minipage}

\bigskip
\noindent See also, for instance, M.\;Spivak [35: p. 140]. The
above phraseology of Riemann may certainly be considered, as
further supporting the aspect (cf. also (1.1) in the preceding)
that the

\bigskip
\noindent (1.4) \hfill
\begin{minipage} {11cm}
    {\it ``space''} consists actually of the objects that fill
    {\it ``it''} up, so that the ({\it differential}\,) {\it
    geometry} on it virtually {\it refers to these} same {\it
    objects}\,; in other words, to the {\it interrelations} among, or
    {\it evolutions} of, them.
\end{minipage}

\bigskip
\noindent Indeed, this very type of ({\it differential}\,) {\it
geometry}, as initiated by K.F.\;Gauss and then extended by
B.G.\;Riemann to any $n$-dimensional {\it ``manifold''}
(:\;Mannigfaltigkeit), {\it concerns} again {\it the object} ({\it
manifold}\,) {\it itself, irrespectively of any surrounding
``space''}. The above dictum of Riemann, may also be related, of
course, with the {\it ``gauge principle''} (cf. (0.2)), the latter
providing thus still another support of (1.4), or even of (1.2),
as above.

Moreover, as further potential applications of ADG, and in
conjunction with the above type of {\it ``generalized
functions''}, \`{a} la Rosinger, we can still mention, as already
hinted at in the preceding, the nowadays {\it ``gauge theories''},
being (F.M.\,Atiyah) ``physical theories of a geometrical
character'', like {\it Yang-Mills theory}, for instance, as well
as {\it geometric (pre)quantization}.

\bigskip
{\bf 2.} Now, by coming back to our previous comments, pertaining
to {\it Leibniz's ``geometric calculus''} (or {\it ``ars
combinatoria''} in his own words, something that he was still
attributing to Analysis too, see the preceding Section 1), we can
further remark that, according to Leibniz,

\bigskip
\noindent (2.1) \hfill
\begin{minipage} {11cm}
{\it ``geometric objects'' do} or, at least, should {\it exist, by
themselves}, independently of any supporting or surrounding {\it
``space''}, the latter being thus simply used to provide us with
the corresponding coordinates.
\end{minipage}

\bigskip
\noindent Consequently, one has to find, as already discussed in
the foregoing, a {\it mechanism} (alias {\it ``calculus''}),
pertaining directly to such objects, without the intervening of
any space, providing the {\it coordinates}, that is, to say, {\it
``location of the objects in the space''}. Of course, even then we
may resort to a certain {\it ``reference point''}, that, however,
finally disappears from our conclusions (equations), something
like, for instance, we effectuate in {\it affine geometry}. This
reminds us here of the famous apostrophe of Archimedes, which goes

\bigskip
\noindent (2.2) \hfill
\begin{minipage} {11cm}
    ``$\mathrm{\Delta \acute{o}\varsigma}$ $\mathrm{\mu o\acute{\iota}}$ $\mathrm{\pi\tilde{\alpha}}$
    $\mathrm{\sigma\tau\tilde{\omega}}$
    $\mathrm{\kappa\alpha\acute{\iota}}$ $\mathrm{\tau\acute{\alpha}\nu}$ $\mathrm{\gamma\tilde{\alpha}\nu}$
    $\mathrm{\kappa\iota\nu\acute{\eta}\sigma\omega}$''
    (:\;give me somewhere to stand and I shall move the earth).
\end{minipage}

\bigskip \noindent Cf. Simplicius [34]. We may still understand,
in that very same way, the r\^{o}le of the {\it ``base-space''}
$X$, used in {\it sheaf theory}, as applied in ADG (see also (2.6)
in the sequel). Hence, by considering here our experience of
today, we can certainly anticipate that such a perspective on {\it
``differential geometry''} should naturally have potential
applications in the r\'{e}gime of nowadays {\it quantum theory},
yet, in particular, in {\it quantum relativity}.

Now, to be fair, we still notice herewith, that the so-called
today {\it ``coordinate-free'' differential geometry} did exactly
the aforesaid job, already (!), concerning the entanglement of
coordinates in our calculations (arguments), referring to
(differenti\-al-)geometric questions, so that, finally, our {\it
conclusions} (formulas) being possible to be {\it stated in a
coordinate-free manner}\,; however,

\bigskip
\noindent (2.3) \hfill
\begin{minipage}{11cm}
the problem was still with the (algebra of) functions, used to do
the job. As we shall perceive, through the subsequent discussion,
the same job could (and, indeed, {\it can}\,) be done by other,
{\it more convenient}, means! (See e.g.\;(0.4), along with the
comments fol\-lowing (2.8) in the sequel).
\end{minipage}

\bigskip
\noindent Thus, classically speaking, one employs here again {\it
smooth functions} on a {\it smooth manifold}, hence, the
appearance of {\it ``singularities''}, where the functions
involved loose their meaning (viz. their calculational power),
notwithstanding, {\it the (in\-trin\-sic\,!) mechanism} of the
method applied (:\;differential geometry) {\it is} still {\it
present, very likely of help} (!), according to our classical (in
the absence of ``singularities'') previous experience, but, {\it
we cannot read it}, due to our apparatus employed (smooth
functions!).\footnote{In this connection, we can further refer to
A.\,Einstein, saying (ibid., p.\;165) that; {\it ``... we can\-not
judge in what manner and how strongly the exclusion of
singularities reduces the manifold of solutions''} [the emphasis
here is ours], viz. the amount of information, we get out from
(or, which is included in) there (i.e., in the (set of)
``singularities'', cf., for instance, Finkelstein, as before).}

Of course, the aforementioned development of recent (classical)
differential geometry (viz. that one on {\it smooth manifolds})
greatly supported and contributed in the formulation of {\it
abstract} (viz. axiomatic) {\it differential geometry}, while it
was underscored and further brought on the stage, by ADG. Thus,
what amounts to the same thing,

\bigskip
\noindent (2.4) \hfill
\begin{minipage} {11cm}
in that {\it ``coordinate-free'' treatment of} CDG it was
virtually hidden the {\it overall power of the inherent mechanism
of (classical) differential geometry}.
\end{minipage}

\bigskip
\noindent That is, in a sense, the much sought after, already by
Leibniz, {\it ``geometric calculus''}. Indeed, and this is the
most {\it fundamental moral}, which one gets out {\it from} ADG;
namely, the fact that:

\bigskip
\noindent (2.5) \hfill
\begin{minipage} {11cm}
{\it that inherent powerful mechanism of} CDG {\it is}, in point
of fact, {\it independent of any notion of Calculus}, in the
classical sense of differential analysis, {\it hence}, and this is
here still worth mentioning, {\it of any notion of smooth
manifold} (providing virtually that Calculus, see (2.6) below),
whatsoever\;!
\end{minipage}

\bigskip
\noindent Consequently, {\it within the same context}, we further
remark that,

\bigskip
\noindent (2.6) \hfill
\begin{minipage} {11cm}
{\it the only contribution}, within the preceding framework, of
the smooth surrounding space was simply to supply the
corresponding (smooth) Calculus, thus, in turn, {\it the smooth
algebra of coefficients}, along with the pertinent, very
instrumental, indeed, {\it ``de Rham exact sequence''}. (For
brevity's sake, we {\it abuse} here {\it terminology}, just
hinting at the (exact) {\it resolution}, in effect, which is
provided, by the corresponding {\it de Rham complex} of the usual
differential forms, as an outcome of the classical, for the case
at issue, {\it Poincar\'{e}'s Lemma}).
\end{minipage}

\bigskip
\noindent Indeed, the next really fundamental inference, one gets
out from the abstract \linebreak (:\;axiomatic) treatment of
differential geometry is that,

\bigskip
\noindent (2.7) \hfill
\begin{minipage} {11cm}
{\it the aforesaid inherent mechanism (``geometric calculus'',}
\`{a} la Leibniz) of CDG {\it is} absolutely {\it rooted on} the
type of {\it ``generalized numbers''}, thus, in the case of CDG,
on the {\it algebra} (in effect, on the {\it algebra sheaf) of}
scalar-valued {\it differentiable} (: smooth) {\it functions},
along with the concomitant ({\it ``exact''}, cf.\,(2.6))
differential {\it de Rham complex}.
\end{minipage}

\bigskip
Thus, by referring to {\it ``Differential Geometry''}, in the
classical sense (of $\cc^{\infty}$-manifolds), one actually means
the study of the structure of a {\it ``locally Euclidean space''},
through the {\it ``geometric calculus''} (in the sense of Leibniz,
as above); that is, to say, in terms of the {\it ``differential
geometric mechanism''} (we might call it ADG), being, anyway,
independent of the particular ``space'' at issue, the same being
virtually based, for the case under consideration, on the
classical {\it ``de Rham differential triad''} (cf., for instance,
concerning the last term, [VS: Chapt. X, p. 278; (1.1)]).

Therefore, in that sense, we can very likely say that this type of
differential geometry cannot be applied in a {\it ``true quantum
gravity''} (see, for instance, C.J.\;Isham [15:\;p.\;400]),
however, {\it by no means} (just, because of the ``space'') its
mechanism, as well; cf. also the ensuing remarks in the present
section, in particular, Section 5 in the sequel.

In point of fact, in contrast to the above, and in full
generality, one realizes, as a moral of ADG, that;

\bigskip
\noindent (2.8) \hfill
\begin{minipage} {11cm}
in the general (:\;abstract) case, {\it any appropriate algebra
(sheaf)}, not even a functional one (!), that is still accompanied
{\it with} a suitable {\it (``exact differential'') de Rham
complex}, can do the same job (: ADG).
\end{minipage}

\bigskip
\noindent As a result, we come thus to the conclusion that,

\bigskip
\noindent (2.9) \hfill
\begin{minipage} {11cm}
{\it it would be}, of course, {\it of paramount importance, any
time we could afford a ``mechanism''}, hence, at the very end, an
{\it ``algebra of coefficients'', incorporating previously
appeared disturbances} (: ``singularities''), {\it being},
however, {\it still able to provide the pertinent ``differential''
set-up!}
\end{minipage}

\bigskip
\noindent As already said in the preceding, the previous data, as,
for instance, in (2.9), are exactly provided by {\it Rosinger's
algebra sheaf}, incorporating, lately, the so-called {\it
``multi-foam algebras''}. This certainly constitutes, so far, an
{\it extremely non-trivial corroboration} of the abstract method,
being, moreover, quite sensible to {\it ``analytic''} questions,
in the classical sense, e.g.\:{\it applications in} PDEs. Yet, it
is a hunch that, very likely, the same abstract method, as above,
will have further potential applications in problems connected
with {\it quantum gravity}. Thus, see e.g. [24; 25], as well as,
the ensuing Sections 3 and 4 below.

\bigskip
{\bf 3.} By commenting further upon our previous argument in
Section 1 (see, for instance, (1.1)), we can still refer to some
relevant thoughts of V.I.\;Denisov and A.A.\;Logunov [2], where
they remark that;

\bigskip
\noindent (3.1) \hfill
\begin{minipage} {11cm}
``Minkowski was the first to discover that the space-time, in
which all physical processes occur, is unified and has a
pseudo-Euclidean geometry. Subsequent study of strong,
electromagnetic, and weak interactions has demonstrated that the
{\it pseudo-Euclidean geometry is inherent in the fields}
associated with these interactions'' [the underline here is ours].
\end{minipage}

\bigskip \noindent
Yet, they also remark that,

\bigskip
\noindent (3.2) \hfill \begin{minipage} {11cm} ``... for an
equation to be covariant it is necessary that it is transformed
according to a tensor law for any arbitrary, admissible coordinate
transformation''.
\end{minipage}

\bigskip
\noindent The same authors attribute the above to V.A.\;Fock [10].
Now, according to the relevant set-up of ADG,

\bigskip
\noindent (3.3) \hfill
\begin{minipage} {11cm}
{\it ``equations'' are expressed} by (sections of) sheaf
morphisms, in effect, {\it by morphisms of vector sheaves}, hence,
in other words, {\it by ``${\mathcal{A}}$-morphisms''} (where
``${\mathcal{A}}$'' stands here for the {\it ``structural}
(algebra) {\it sheaf''}, alias {\it ``generalized numbers''} of
the theory). Therefore, by their very definition, in terms of {\it
tensorial morphisms}$\;$(!), hence, in accord with the so-called
{\it ``principle of general covariance''}; see, for instance, {\it
Yang-Mills equations}, in terms of the curvature, still, within
the framework of ADG. Cf. [17: p. 167; (2.3)], or even [21: Vol.
II; Chapt. VI, (4.76)/(4.77)].
\end{minipage}

\bigskip
\noindent This, of course, constitutes further another vindication
of the {\it naturalness of} ADG (viz., in effect, of the {\it
sheaf-theoretic} treatment of the same). It seems that {\it
everything is inherent} there, alias {\it ``innate''} (I owe the
latter suggestive (synonymous) expression to I.\;Raptis).

Within the same vein of ideas concerning (3.1), about the same
time, T.H.\;Parker [29], remarks that,

\bigskip
\noindent (3.4) \hfill
\begin{minipage} {11cm}
{\it ``... the topology is inherent in the field''} (!);
\end{minipage}

\bigskip
\noindent still the exclamation sign here is ours. Yet, we also
remark that the notion of {\it topology in field theory} is, in
point of fact, a matter of {\it homotopy}, and at the very end, of
{\it algebraic topology}, as e.g. cohomology classes, Poincar\'{e}
Lemma, de Rham complex, characteristic classes and the like. Thus,
finally, we may say that:

\bigskip \medskip
\noindent (3.5) \hfill
\begin{minipage} {11cm}
the notion of {\it ``field''} seems to be herewith predominant and
overwhelming, being further inextricably connected with that one
of an ``(ele\-mentary) {\it particle}''. Indeed, according to the
technical part of ADG,

\medskip
\noindent (3.5.1) \hfill
\begin{minipage} {9cm}
{\it the two notions}, as above, {\it may be viewed, as
identical}.
\end{minipage}
\end{minipage}

\bigskip \medskip
\noindent In this concern, the notion of {\it ``field''} appears
thus, as a fundamental one, and, as in the classical case, {\it
``not further reducible''} (A.~Einstein); on the other hand,
within the context of ADG, this now is {\it independent of any
``surrounding space''}, while we are still able to {\it employ
directly on the ``fields''} the whole machinery of ADG, to the
extent, at least, that this is feasible, thus far.

Furthermore, the above deliver us from the classical {\it
``drawback that the continuum brings''} (A.~Einstein, again). As a
matter of fact, we are trapped here into the latter notion, as a
result, in effect, of our adherence to the concept of {\it
``space-time continuum''}, as an appropriate
$({\mathcal{C}}^\infty -)${\it manifold}. However, as already
pointed out in the preceding (cf., for instance, (2.4)), this is
{\it no more necessary, the machinery of} ADG {\it being still in
force}, without it; in other words, we can still say (see also,
for instance, (0.3) in the preceding), that,

\bigskip
\noindent (3.6) \hfill
\begin{minipage} {11cm}
we are thus able to {\it ``formulate statements about a
discontinuum without calling upon a continuum space-time'',}
\end{minipage}

\bigskip
\noindent something that also provides us with the possibility of
thinking of the {\it ``real''}, without the need of resorting,
inevitably, to the {\it ``continuum''}, a disputable, at the very
end, point of view, apart, of course, from its own (mathematical)
definition.

Moreover, within the same context, we can still refer here, once
more, to A.~Einstein himself, by saying that (emphasizing is
ours);

\bigskip
\noindent (3.7) \hfill
\begin{minipage} {11cm}
{\it ``... Adhering to the continuum originates} with me {\it not
in a prejudice}, but arises out of the fact that I have been {\it
unable to think up anything organic to take its place ...''}
\end{minipage}

\bigskip
\noindent See A.~Einstein [5: Vol. 2, p. 686]. Yet, we can also
refer at this place to R.P.~Feynman's criticism, by referring to
the {\it ``continuum in the quantum deep''}, thus saying, among
other things, that;

\bigskip
\noindent (3.8) \hfill
\begin{minipage} {11cm}
``...\,the theory that space is continuous is wrong, because we
get ... infinities [viz.\;{\it ``singularities''}\,] and other
similar difficulties ... [while] the simple ideas of geometry,
extended down to infinitely small are wrong!''
\end{minipage}

\bigskip
\noindent Consequently, we thus  realize here too, either the lack
of an {\it ``organic theory''}, which would be able to cope with
the absence of an a\;\;p\;r\;i\;o\;r\;i ``continuum'', or even,
and more so, with the occasional appearances of infinities.
However, we can still remind us here that, avoidance of {\it
``infinities''} (:\;{\it ``something too great''}) does not
pertain, anyhow, to {\it ``sensible mathematics''}. That is,

\bigskip
\noindent (3.9) \hfill
\begin{minipage} {11cm}
{\it ``Sensible mathematics involves neglecting a quantity when it
turns out to be small~$-$~not neglecting it just because it is
infinitely great and you do not want it''.}
\end{minipage}

\bigskip
\noindent See P.A.M.~Dirac [3: p. 36].

Yet, when referring, just before (cf. (3.8)), to ``simple ideas of
{\it geometry}'', one rather means those (simple, viz.
fundamental) principles of {\it ``differential geometry''}, that
we wanted to be applicable in the quantum deep, as well. But, as
already hinted at in the preceding,

\bigskip
\noindent (3.10) \hfill
\begin{minipage} {11cm}
the power and effectiveness of {\it ``differential geometry''}
rests, in effect, in its {\it inherent (:\;innate) mechanism},
being of an algebraic (:~operational) character, independently of
any surrounding space, the same mechanism referring, in point of
fact, directly, to the {\it ``objects''}, we are dealing with.
\end{minipage}

\bigskip
\noindent In this context, and in close connection with our last
comments in (3.10) above, which further point to the {\it aptitude
of} ADG {\it for confronting with} recent perspectives on {\it the
subject}, concerning the correspondence,

\bigskip
\noindent (3.11) \hfill
\begin{minipage} {11cm}
{\it differential geometry} (in effect, its {\it mechanism}) $-$
``space'',
\end{minipage}

\bigskip \noindent
one can also mention Finkelstein's apostrophe, in that,

\bigskip
\noindent (3.12) \hfill
\begin{minipage} {11cm}
{\it ``... we take acts as basic instead of points''.}
\end{minipage}

\bigskip
\noindent See D.R.~Finkelstein [9: p. 425]. Indeed, we can still
say that we have again here, once more, another variant of {\it
``Klein's principle''} (:~``space'' is determined by (the group
of) its automorphisms). Furthermore, as Denisov and Logunov remark
(loc. cit.), see also (1.1) in the preceding,

\bigskip
\noindent (3.13) \hfill
\begin{minipage} {11cm}
``Pseudo-Euclidean space-time is not a priori, i.e., given from
the start, or having an independent existence. It is an integral
part of the existence of matter, ... it is [always] the geometry
by which matter is transformed.''
\end{minipage}

\bigskip
\noindent In this concern, we can also remark here that,

\bigskip
\noindent (3.14) \hfill
\begin{minipage} {11cm}
{\it matter is transformed, according to the} (dynamics of the)
{\it physical law}.
\end{minipage}

\bigskip
\noindent Within the same point of view, we further note that one
can still understand the classical, {\it ``matter tells space how
to curve''} (cf. [27: p. 5]), exactly in the sense of (3.14), that
is,

\bigskip
\noindent (3.15) \hfill
\begin{minipage} {11cm}
{\it space} (:\;matter) {\it is curved, according to} (the
dynamics of the) {\it physical law}.
\end{minipage}

\bigskip \noindent Yet, by also referring to (3.14), we can
actually say that,

\bigskip
\noindent (3.16) \hfill
\begin{minipage} {11cm}
{\it the variation} (transformation) {\it of the matter is
equivalent with the existence of an $\ca$-connection} (viz.,
physically speaking, with the {\it physical law} itself).
\end{minipage}

\bigskip
\noindent Indeed, by restricting ourselves to a (free) {\it
boson}, for instance, we can associate (3.15) with the basic
relation of ADG,
$$
\delta (\theta_{\alpha})={\tilde \partial}(g_{\alpha \beta}),
\leqno{(\mbox{3.17})}$$
so that our assertion in (3.15) is just a consequence of the
so-called {\it Atiyah's criterion} (for the existence of an
$\ca$-connection: see [VS: Chapt. VII, p. 115]). As a matter of
fact, an analogous formula to (3.16) holds true, for any {\it
vector sheaf} $\ce$, in general; viz. one has
$$\de (\om^{(\al)})=\tilde{\partial}(g_{\al
\be}),\leqno{(\mbox{3.18})}$$
{\it ``transformation law of potentials''}, where one sets
$$\de (\om^{(\al)}):= \om^{(\be)}-Ad(g^{-1}_{\al \be})\om^{(\al)},
\leqno{(\mbox{3.19})}$$
such that
$$(\om^{(\al)})\in C^\circ (\cu ,M_n (\Om
))\leqno{(\mbox{3.20})}$$
stands for the corresponding $\ca$-{\it connection $0$-cochain
matrix form} of $\ce$, associated with a given {\it local frame}
$\cu$ of it: loc. cit., p. 113, Theorem 2.1.

Yet, within the same context, as above one still realizes that;
({\it 3.14}\,) {\it implies}, in effect, ({\it 3.16}\,), in the
sense that:

\bigskip \noindent (3.21)\hfill
\begin{minipage}{11cm}
    the {\it variation of the matter} is actually the way {\it the
    physical law itself} is revealed to us; that is, the {\it
    ``causality''} itself, or else (in technical, viz.
    mathematical, terms) the $\ca$-{\it connection} (differential
    equation), detected, finally, through its {\it strength}
    (:\;curvature/solution of the corresponding differential
    equation at issue).
    See also the same rel. (3.19), as above.
\end{minipage}

\bigskip
Now, looking at the things {\it locally} (hence, what we actually
observe/measure), we can say that, the whole story is virtually
reduced to
$$\ca ut {\mathcal L}, \leqno{(\mbox{3.22})}$$
viz. to the {\it group sheaf of automorphisms of} $\mathcal{L}$,
the carrier (:\;{\it ``space''}) of the (bare) {\it boson} at
issue. Thus, in general, by considering a {\it vector sheaf $\ce$,
of rank} $n\in \bn$, over $X$, carrier of a (bare) {\it fermion}
(see also A.\;Mallios [21:\;Vol.\;I; Chapt. II, (6.29)],
con\-cern\-ing the terminology employed herewith), one looks at
the corresponding {\it group sheaf of automorphisms of} $\ce$,
$$\ca ut \ce .\leqno{(\mbox{3.23})}$$
Therefore, by taking the things {\it locally} (:\;on a {\it local
gauge $U$ of}\; $\ce$), one gets;
\[
\begin{aligned}
    (\ca ut \ce )\big|_{U} &=\ca ut (\ce \big|_{U})=\ca ut (\ca^n
    \big|_{U})=\cg \cl (n, \ca \big|_{U})\\
    &=(\ca ut \ca^n)\big|_U = \cg
    \cl (n, \ca)\big|_U ,
\end{aligned}
\leqno{(\mbox{3.24})}
\]
so that one still obtains,
\[
    (\ca ut \ce )(U)=Aut(\ce \big|_{U})=\cg \cl (n, \ca) (U) =GL(n,\ca
    (U)).\leqno{(\mbox{3.25})}
\]
See also loc. cit., Chapt. I; Section 6, in particular, (6.31) and
(6.36).

Thus, in view of (3.22) and (3.23), as above, one further has
herewith another instance of {\it Klein's point of view}, as
before, so that one effectuates, once more (see also (1.1) in the
preceding), that, in our case,
$$\mbox{\it ``space'' is} \ \ \ce ,\leqno{(\mbox{3.26})}$$
in point of fact, {\it it is represented by} $\ce$, or even, {\it
equivalently}, according to the Kleinian perspective, as before,
{\it by its} ({\it group sheaf of}) {\it automorphisms}, $\ca ut
(\ce )$. This latter aspect might also be related, pretty well,
with nowadays tendencies to look, namely, at the {\it ``quantum
deep''}, as something of a {\it foamy, fuzzy}, and the like,
nature (cf., for instance, {\it ``quantum soup''}(!)).
Furthermore, the same point of view, as well, may be connected
when arguing, at least, within a {\it ``vacuum''}, with a quite
recent theoresis of the above, in terms of {\it ``quantum causal
sets''}; see thus, for instance, I.\,Raptis [31]. In this
connection, we can still remark that the relevant here {\it
``inner product problem''} and the {\it ``problem of time''}
(ibid.), are naturally transferred to the respective {\it
sheaves}, that, within ADG represent the {\it objects} (e.g.
elementary particles) at issue, in particular to the {\it group
sheaf of automorphism}, $\ca ut \ce$, as above, this being, in
effect, through its {\it action}, the {\it ``dynamics'' itself of}
the corresponding (vector sheaf) $\ce$ (:\;``problem of time''\,).
Yet, by defining an {\it ``inner product'' on} $\ce$, we still
require this to be {\it unaltered, under the action of} $\ca ut
\ce$ (:\;``inner product problem''\,). So the previous two
fundamental problems of the standard theory referred, in that
classical context, to ${\mathcal{D}}${\it iff} $(X)$, group of
automorphisms (:\;diffeomorphisms) of the base (space-time)
manifold $X$, hence the occasionally concomitant ``singularities''
(!), {\it are} thus {\it reduced} here {\it to} $\ce$, therefore,
{\it locally to} $\ca$, our {\it ``arithmetics''}, which, as
already said, {\it can} just now {\it incorporate
singularities}\,; see also (3.28) below, as well as, (6.1) in the
sequel, or even A.\;Mallios [21: Vol. II; Chapt. IX].

Yet, by still commenting on the above {\it ``Kleinian
perspective''}, we can further set out the following meditations
(argument):

\bigskip \noindent (3.27) \hfill
\begin{minipage}{11cm}
    {\it Evolution} is an algebraic automorphism (Feynman),
    expressed analytically, via the Hamiltonian (Schr\"{o}dinger,
    viz. $\partial_t =H$, {\it ``evolution''}\,); this, in turn,
    entails again, still algebraically
    (Heisenberg-Prigogine-K\"{a}hler-Hiley), the {\it ``time
    operator''} (:\;description, in time, of the physical system).
\end{minipage}

\bigskip \noindent
The previous conclusion has been motivated mainly from a quite
recent account thereof of B.I.\;Hiley [14]. (I am indebted to
Prof.\;Hiley for kindly giving to me access to his relevant
manuscript). Still, the {\it sheaf-theoretic character''} of
(3.23) may also be construed, as being in accord with the above
{\it ``evolutionary''} point of view, advocated by (3.27).

On the other hand, the previous aspect, concerning $\ca ut (\ce
)$, still falls in with the standard {\it ``locality principle''}
(see also (3.24), along with (3.29) below, or even [22: p. 1896]).
That is, in other words,

\bigskip \noindent (3.28)\hfill \begin{minipage}{11cm}
the same {\it ``principle of locality''} stands in complete
agreement with the {\it sheaf-theoretic} flavor of the present
treatment, in that, both aspects lead naturally {\it from local
information to global perspective}, while the very notion of a
sheaf is still quite akin to the {\it ``relativistic''}
(:\,varying) {\it aspect}, yet, {\it equivalently}, in terms of
{\it sections} (:\,our calculations\,!).
\end{minipage}

\bigskip
On the other hand, the preceding point out, once more, to the {\it
significance}, as well, of the {\it ``structure sheaf}, or else
{\it ``sheaf of coefficients''}, $\ca$, for the whole subject, in
such a manner that, roughly speaking, one can say that;

\bigskip
\noindent (3.29) \hfill
\begin{minipage} {11cm}
{\it everything} (concerning our calculations) is {\it locally}
reduced to $\ca$, {\it equivalently}, to {\it sections} of the
same sheaf.
\end{minipage}

\bigskip \noindent An analogous conclusion is still valid
for the {\it space of} $\ca$-{\it connections of} $\ce$,
$$\cc onn_{A} (\ce ),\leqno{(\mbox{3.30})}$$
hence, for the corresponding {\it moduli space of} $\ce$,
$${\mathcal M}(\ce )\equiv \cc onn_{A}(\ce )/\ca ut \ce
,\leqno{(\mbox{3.31})}$$
as well, where too the things are locally reduced to $n\times n$
{\it matrices} (of {\it sections}) {\it of} $\ca$, and/or similar
ones of (sections of) {\it ``$1$-forms''}, viz. of the $\ca$-{\it
module} $\Om^1$, {\it hence}, finally, {\it of} $\ca$ again, if,
as is usually the case, $\Om^1$ is still a {\it vector sheaf} on
$X$. (In this regard, cf. also [VS: Chapts VI, VII]).

Of course, once more, it goes always without saying that,

\bigskip
\noindent (3.32) \hfill
\begin{minipage} {11cm}
{\it we gain} very much {\it in insight, any time we} are able to
{\it free our conception}, we have about a specific physical
problem, {\it from any} reference to some particular {\it
coordinate system (:\;``space'')}.
\end{minipage}

\bigskip
\noindent In this context, we can also remark that

\bigskip
\noindent (3.33) \hfill
\begin{minipage} {11cm}
{\it ``local gauges''} (viz.~{\it ``coordinates''}, or even {\it
``space''}) are used here for our calculations, yet, to detect the
(independently of the former existing) {\it physical law}, while
to understand the latter one has to free his conception of the
former (viz. of the means applied).
\end{minipage}

\bigskip
\noindent So, within the same vein of ideas, and, technically
speaking, as it concerns the operational (alias, litourgical) part
of ADG, we can further say that,

\bigskip
\noindent (3.34) \hfill
\begin{minipage} {11cm}
{\it algebraic topology} can be conceived, as a {\it ``relational}
hence, algebraic, in nature {\it way of looking at the topology}
[:~space]'', therefore, as {\it more akin to physics}.
\end{minipage}

\bigskip
\noindent Thus, we come still here to the relevant point of view,
emphasized by C.~von~Westenholz [39: p. 323], when saying that;

\bigskip
\noindent (3.35) \hfill
\begin{minipage} {11cm}
``The mathematical structure underlying field quantities ... is
essentially de Rham cohomology''.
\end{minipage}

\bigskip
{\bf 4.} Accordingly, the problem, as well as, our own progress in
confronting with it, is thus, very likely, lying, to quote here
W.\;Stevens [36: p. 184], in a {\it ``movement through changes in
terminology''}, which we finally apply, when look at the
particular problem, we are interested in.

We wish to terminate the previous discussion by just referring, as
an example of the viability of the preceding thoughts, to other
authors of the very recent past (the {\it ``Polish school''}
mentioned in the above, see Section 0), in whose work one also
finds similar considerations to the foregoing material; thus, we
read, for instance, the utterance that,

\bigskip
\noindent (4.1) \hfill
\begin{minipage} {11cm}
{\it ``... the imagination is very often restricted by constraints
coming from the language we use''.}
\end{minipage}

\bigskip
\noindent See M.\;Heller et al. [13: p. 54]. Yet, we still recall,
in that context, Einstein's own motto that, {\it ``imagination is
more important than knowledge''}. So, everything that could affect
our imagination (the language we use, for example), affects, in
point of fact, our ability of knowing (:\;describing) the reality.

Finally, as an overall moral of the preceding, we still conclude
the following two-way (:\;``amphidromous'') relation,

\bigskip
\noindent (4.2) \hfill
\begin{minipage} {10cm}
\begin{picture}(50,0)
\put(0,-1.8) {\makebox(0,0){{\it physics}}}%
\put(75,0) {\vector(-1,0){50}}%
\put(75,0) {\makebox(0,0)[r]{\vector(1,0){50}}}%
\put(140,-1.8) {\makebox(0,0){({\it differential}) {\it
geometry},}}
\end{picture}
\end{minipage}

\bigskip
\noindent the second member of the above diagram hinting actually
at the {\it innate mechanism}, viz.\;the {\it ``ars
combinatoria''}, or even {\it ``geometric calculus''}, \`{a} la
Leibniz, of the mathematical discipline at issue, as explained in
the foregoing. Accordingly, by simply paraphrasing herewith S. Mac
Lane [16: p. 257], we realize that;

\bigskip
\noindent (4.3) \hfill
\begin{minipage} {11cm}
``{\it geometry} is not just a subdivision or a subset within
Mathematics, but {\it a means} ...'' {\it to get out of phenomena}
(:\:physics) {\it formal rules} (:\:physical laws),
\end{minipage}

\bigskip
\noindent that also fits in well with the preceding, justifying
further (4.2). In that sense a formal rule, as before, that is, in
other words, a {\it physical law} cannot be dependent on (or even
restricted by) a {\it ``singularity''}. Yet, to paraphrase here
A.~Einstein [6: p. 164], {\it ``singularities must be excluded''}
from a procedure, whose function can be described, according to
(4.3).

Thus, what we actually perceive, as {\it ``laws of Physis''} is
given, in the sense that these ``laws'' are there. On the other
hand, it is {\it we}, who do not provide the proper theories to
follow (:\:``understand'', or even better, to {\it describe})
these laws, yet to further predict their evolution. So it is in
that very sense that;

\bigskip
\noindent (4.4) \hfill
\begin{minipage} {11cm}
``{\it the ``laws'' of Nature cannot have} anomalies, alias {\it
``singularities''}, an attribute that is virtually ours, since
here again, {\it the manner in which} these {\it laws function
is}, certainly, still, {\it given}\,!
\end{minipage}

\bigskip \noindent
Finally, within this same vein of ideas, we can further remark
that;

\bigskip
\noindent (4.5) \hfill
\begin{minipage} {11cm}
{\it we are virtually part of} what we understand as {\it Nature,
not the creators} of the latter. Thus, what we ascribe, as
anomalies, {\it ``singularities''} etc., are just {\it our own
verdict} (interpretation) about Nature and {\it not ``particular
instances''} of {\it it}.
\end{minipage}

\bigskip
{\bf 5.} Now, let us come back again to the aforementioned remarks
of R.P.\;Feynman (see (3.8)), according to which,

\bigskip
\noindent (5.1) \hfill
\begin{minipage} {11cm}
{\it ``the simple ideas of geometry, extended down to infinitely
small are wrong}\,!
\end{minipage}

\bigskip \noindent
Thus, by what has been said in the preceding, concerning the above
apostrophe of Feynman, one may remark, that we can virtually have
here a {\it quite different situation}, pertaining, at least, to
the {\it differential geometry}, in point of fact, to {\it its
inherent} (differential-geometric) {\it mechanism}, yet, in other
words, the implemented thereof {\it ``ars combinatoria''} \`{a} la
Leibniz. That is, although we cannot speak of the space (roughly
speaking, the {\it ``continuum''}) that entails, alias supports,
the {\it standard differential geometry of smooth manifolds},
notwithstanding, we can still apply, even to that extended deep
the accompanied (innate) mechanism of that geometry, that is, in
other words, its {\it differentiable functions''}, suitably {\it
generalized, along with the} attached to them {\it
differential-geometric technique} that still seems to work, even
in {\it that r\'{e}gime}\,!; indeed, this is actually due, to the
very nature of the generalized functions (in effect, {\it
sections} of appropriate sheaves), which, thus, are involved.

So, in this context, we can virtually assert that,

\bigskip
\noindent (5.2) \hfill
\begin{minipage} {11cm}
{\it ``the simple ideas of} [differential] {\it geometry''}, [at
least], {\it are not wrong}\,!
\end{minipage}

\bigskip \noindent
Indeed, as already explained in the foregoing, the essence (:
inherent mechanism) of \linebreak (differential) geometry may
still be applied in that deep, in spite of the presence of an
extremely {\it ``anomalous''}, in the classical sense, carrying
space (if any\,!). Yet, once more, the impediment here is with the
classical notion of the supporting space, providing, in effect,
the standard {\it ``differential triad''}, and {\it not} with the
latter p\;e\;r\;\;s\;e, which is virtually independent of the
former, that in the quantum deep {\it seems} (!) {\it to be}
different from the usual one. Presumably, what was for Feynman in
conflict with the geometry in the quantum domain, was the way of
applying it therein; accordingly, to paraphrase him, we can say
that,

\bigskip \noindent (5.3)\hfill \begin{minipage}{11cm}
if we are going to apply any {\it ``geometrical reasoning''}
within the quantum context, this should be done {\it not in a
geometrical} (classical) {\it way} (!), {\it but in an analytic}
(algebraic) {\it one with symbols}.
\end{minipage}

\bigskip \noindent See R.P.\;Feynman [7: begin of p. 44]. Now, it is, actually,

\bigskip \noindent (5.4)\hfill \begin{minipage}{11cm}
{\it this analytic $($operational-theoretic$)$ character}, that
{\it permeates} the {\it classical $($``coordinate-free''$)$
differential geometry}, that has been brought on the stage, quite
perspicuously, indeed, by ADG, showing thus that {\it the former
is}, in effect, {\it independent of any surrounding ``space''}.
\end{minipage}

\bigskip \noindent Hence, the potential applicability of this
machinery to {\it quantum gravity}.

Accordingly, and expressed differently, the apparent disagreement
of {\it general relativity} with {\it quantum theory}, the former
being rooted on classical differential geometry of the so-called
{\it ``smooth''} manifolds, therefore, the difficulties, as well,
when trying to apply the latter to problems in {\it quantum
gravity}, appear, as a result of the preceding, to be due, very
likely, {\it not} thus much to the mechanism (:\;idea)
p\;e\;r\;\;s\;e of differential geometry, as to the type of the
same, that is, to say, to the corresponding sort of {\it
``differential triad''}, in the sense of ADG, that we usually
employ in that context!

That is, to put it yet in another way, the fault is with the way
we understand resulting the {\it classical differential-geometric
mechanism}; thus, in that point of view, this mechanism is just an
outcome of the standard notion of a (differential-smooth)
manifold. Nevertheless, this is simply a misinterpretation, while
the essence of the matter is much more deeper, referring, in
effect, to an extremely inherent feature of this mechanism, being,
in principle, independent of any intervening space. Indeed, one
can still apply the {\it abstract differential-geometric
technique} even in much more general situations, see e.g. [22],
[23], [24], [25].

\bigskip
{\bf 6.} On the other hand, by looking at potential physical
applications of ADG, the base space of the sheaves involved does
not necessarily have the properties we usually ascribe to the
analogous ``base space'' (:\;space-time) of the classical theory.
So, by contrast with what we usually do classically, in a {\it
sheaf-theoretic} treatment of the situation, {\it in terms of}
ADG, by considering the properties we want to (or, at least, feel
that we should) have,

\bigskip
\noindent (6.1)\hfill
\begin{minipage} {11cm}
{\it we virtually transfer all the} desired {\it properties}, as
above, {\it to} (the stalks of) {\it the sheaves}, that, according
to {\it our} theory, for that matter, {\it represent the objects},
we wish to study (as, for instance, elementary particles).\;Of
course, this is also in accord with our present-day conception of
{\it physics}, the latter being actually concerned with relations
rather (i.e., equations/laws), yet, with functions/sections, that
``live'' on a given ``space'', than this ``space'' itself (!),
whose existence  p\;e\;r\;\;s\;e is, for that matter, quite
disputable, anyway.
\end{minipage}

\bigskip \noindent
Precisely speaking, all the desired properties, that are
classically addressed to the underlying (space-time) manifold, are
here, that is, within the {\it framework of} ADG, transferred to
the corresponding {\it ``arithmetics''}, alias {\it ``sheaf of
coefficients''} $\ca$, and finally, in an appropriate manner (here
{\it paracompactness} of the underlying topological space, base
space of the sheaves involved, and {\it fineness of} $\ca$ are
applied), to the $\ca$-{\it modules} (yet, {\it vector sheaves}\,)
concerned.

In this connection, we can still say that, as a {\it general moral
of} nowadays {\it quantum theory}, we usually try to

\bigskip \noindent (6.2)\hfill
\begin{minipage}{11cm}
    {\it abandon ``space''} and look, instead, at the objects which
    ``live'' on the space and their interrelations, {\it directly},
    viz. without referring and/or being influenced by ``properties''
    of the space.
\end{minipage}

\bigskip \noindent
Therefore, in that perspective,

\bigskip
\noindent (6.3)\hfill
\begin{minipage} {11cm}
{\it we} can further {\it apply} the mechanism of (abstract) {\it
differential geometry to the} pertinent {\it objects},
p\;e\;r\;\;s\;e, without thus referring to any supporting space,
exactly, because ADG {\it does not depend} on the latter.
\end{minipage}

\bigskip \noindent
Plus, the very nature of the {\it ``arithmetics''} applied (!),
according to which, we realize that,

\bigskip \noindent (6.4)\hfill \begin{minipage}{11cm}
based further on ADG, we are thus no more compelled to find {\it
``solutions free of singularities''} (A.\;Einstein [6:\;p.\;165]),
but (to af\-ford that particular {\it ``arithmetics''}, so that to
be able) to state {\it ``equations''}, whose solutions can engulf
the {\it ``singularities''}.
\end{minipage}

\bigskip \noindent In this connection, one can consider the
aforementioned {\it ``Eddington-Finkelstein coordinates''},
simply, as an anticipating example of the above, as already hinted
at by (0.6) in the preceding. Now, the same is still suggesting
that we might have just here a machinery, suitable to a {\it
``relativistic quantum mechanics''}, to quote also P.A.M.\;Dirac
[4:\;p.\;85], {\it ``in which we will not have ... infinities
occurring at all''} [the emphasis is ours]. Thus, in view of the
foregoing, this will be the new r\^{o}le of the {\it
``arithmetics} (in point of fact, of the appropriate {\it
``differential triad''}), we would choose, each time, depending on
the particular problem at issue.

Finally, by further looking at the two {\it fundamental
principles} of nowadays physics, as stated in (0.1) and (0.2) at
the beginning of the present discussion, we still remark that even
by the very formulation of these two principles, we implicitly
assume in effect that the r\^{o}le of the {\it ``space''},
involved therein is quite a secondary one (if any, at all!). {\it
This}, somehow obscured situation due, in point of fact, to the
way we were employing hitherto the (classical) differential
geometry, {\it is} well {\it pointed out}, exactly, {\it by} the
whole technique of {\it Abstract Differential Geometry}, as this
has been succinctly indicated in the preceding discussion.

Yet, we can still say here that the {\it potential value} of the
above perspective, as this is exhibited by ADG, lies exactly in
its {\it generality} and {\it presentation of general fundamental
principles}, that actually govern CDG (which may thus be regarded
as a `subtheory' of ADG); we are thus, in that sense, in accord
with a relevant aphorism of $G$. Darboux, saying that (the
emphasis below is ours);

\bigskip \noindent
(6.5)\hfill \begin{minipage}{11cm}
    {\it ``Le caract\`{e}re propre des m\'{e}thodes ... consiste
    dans l' emploi d' un petit nombre de principes
    g\'{e}n\'{e}raux ...; et les cons\'{e}quences sont d' autant
    plus \'{e}tendues que les principes eux-m\^{e}mes ont plus de
    g\'{e}n\'{e}ralit\'{e}.''}
\end{minipage}

\bigskip \noindent
See G. Darboux [1: Introd., p. 3]
\\

{\bf 7.} We close the present discussion, by considering, as an
example of the preceding, in particular of that part of it,
concerning {\it Rosinger's algebra sheaf}, the well-known {\it
Uhlenbeck's theorem} on {\it ``removable singularities''},
pertaining thus to a (smooth) extension of a {\it Yang-Mills
field}, defined on $S^4$, modulo the north pole (see [38]; yet,
T.H.\;Parker [29] gave a generalization of the same result.

Now, by considering the {\it sheaf of generalized functions} \`{a}
la {\it Rosinger}, we can further look at {\it Uhlenbeck's
Yang-Mills field}, as defined on
$$ S^4 \backslash \{ \infty \} \equiv U \subseteq S^4 \equiv X
\leqno{(\mbox{7.1})}$$
where the {\it closed set} (in effect, a {\it nowhere dense set}
in $S^4$), $\{ \infty \} \subseteq S^4$ (:\;{\it ``north pole''}),
is a {\it ``singularity''} (set), in the classical sense. Thus,
since {\it Rosinger's algebra sheaf} (i.e., our {\it ``structural
sheaf''} $\ca$ is {\it flabby}, (see e.g.
A.\;Mallios-E.E.\;Rosinger [24], and/or [25]), one gets that {\it
the canonical map} (cf. (7.1)),
$$\Ga (X,\ca )\to \Ga (U,\ca ) \leqno{(\mbox{7.2})}$$
{\it is surjective}, by the very definition of the {\it
flabbiness} of $\ca$, so that any (continuous) {\it local} section
of $\ca$ over an {\it open} set $U\subseteq S^4$ is extended to a
(continuous) {\it global} section of $\ca$ over $S^4$.

Therefore, by considering a {\it Yang-Mills field}
$$(\ce ,D)\leqno{(\mbox{7.3})}$$
on $S^4 \equiv X$, the $\ca$-connection $D$ is locally expressed,
via a {\it local gauge}, say,
$$U_0 \subseteq U \leqno{(\mbox{7.4})}$$
of $\ce$, through a so-called {\it local $\ca$-connection matrix
of $D$, associated with} $U_0$,
$$\om \equiv (\om_{ij})\in M_n (\Om (U_0 ))=M_n (\Om )(U_0 ),
\leqno{(\mbox{7.5})}$$
with $1\le i,j \le n \equiv rk\ce$; yet, the $\ca$-{\it module
$\Om$ of $1$-forms} is, for the case at issue, also a {\it vector
sheaf} on $X$, so that
$$\om_{ij}\in \Om (U_0 )\cong \ca (U_0 )^m ,\leqno{(\mbox{7.6})}$$
{\it with} $m=rk\Om$. Therefore, based on (7.2),

\bigskip \noindent (7.7)\hfill \begin{minipage}{11cm}
any {\it continuous (local) section of $\Om$ on} $U_0$, hence, in
view of (7.5), a {\it local realization}, in effect, {\it of $D$
on $U_0$, that is}, actually, {\it of $(\ce ,D)$ itself, can be
extended to the whole space} $S^4$, as well.
\end{minipage}

\bigskip \noindent
The above constitutes, virtually, an {\it ample generalization} of
the classical result of K.\;Uhlenbeck (loc.\;cit.;\;p.\;24,
Theorem 4.1). Yet, it is worth remarking herewith that in our
argument, employed in (7.7),

\bigskip \noindent (7.8)\hfill \begin{minipage}{11cm}
{\it no a p\;r\;i\;o\;r\;i\; restriction on the ``energy''}
(:\,field strength $=$ curvature of $D$), as one assumes in the
classical case (ibid.), is actually required. Moreover, {\it no
restriction on the dimension} of the ``space'' is necessary
either.
\end{minipage}

\bigskip \noindent
On the other hand, it is still worth mentioning here, that the
functions, in point of fact, {\it sections}, which are involved in
the previous account of the classical result under discussion, may
have a far bigger amount of {\it ``singularities''}, that is, to
say, {\it prohibiting places, concerning the standard theory};
see, for instance, A.\;Mallios-E.E.\;Rosinger [24], or even [25].
In conclusion, one thus obtains, within the general setting of ADG
(by contrast, cf. also K.K.\;Uhlenbeck [38: p. 11, Abstract]),
that;

\bigskip \noindent (7.9)\hfill \begin{minipage}{11cm}
one can consider {\it solutions of the Yang-Mills equations} over
any {\it ``space'' with ``singularities''}, that, for instance,
can be {\it engulfed in} a {\it Rosinger multi-foam algebra sheaf}
(cf. [25]), the latter being viewed, as a {\it ``sheaf of
coefficients''}, in the sense of ADG.
\end{minipage}
\bigskip \bigskip
{\bf 8.} {\bf Appendix: Quantum gravity}.$-$ By taking {\it second
quantization} into account (see, for instance, [20] or even [21:
Chapt. II]), one concludes that:

\bigskip \noindent (8.1)\hfill
\begin{minipage}{11cm}
an {\it equation} expressed {\it in terms of sheaf-morphisms}
(take e.g. the {\it Yang-Mills equation}, referring to a given
Yang-Mills field) {\it is}, in effect, by definition, {\it
relativistic, with the field} itself {\it being the variable}.
Furthermore, {\it the same equation} refers, in point of fact,
already, {\it to the quantum} of the field, {\it this still
being}, by the very definitions (:\;{\it second quantization},
again!), just {\it the elementary particle} in terms of which the
equation in focus has been formulated (see also A.\;Mallios [21:
Chapt. VII; (5.8), (5.11)]).
\end{minipage}

\bigskip \noindent
Now, in this context, we further note that;

\bigskip \noindent (8.2)\hfill
\begin{minipage}{11cm}
the presence of the {\it field} (:\;{\it elementary particle}, see
also e.g. loc. cit. (1.5)) in an equation, as before, is virtually
exhibited, through the corresponding {\it field strength} (ibid.
(1.9)). Hence, in terms of a {\it ``structure sheaf invariant''}
(alias, in the terminology employed herewith, by means of an {\it
``$\ca$-invariant''}, or even, classically speaking, in terms of a
{\it ``tensor''}), that is, in other words, via an entirely {\it
physical} notion.
\end{minipage}

\bigskip \noindent
As a result, though it might probably sound a bit strange, and
paraphrasing actually Leibniz, herewith (cf. (2.1) in the
preceding), we can say that;

\bigskip \noindent (8.3)\hfill
\begin{minipage}{11cm}
{\it space-independent notions are} the real {\it physical} ones!
(This might also be related with (1.1)).
\end{minipage}

\bigskip \noindent
So, by saying here {\it ``physical''}, we can also refer to {\it
``geometrical''}, this being actually the connection with Leibniz,
as alluded to above, an aspect that might further be connected
with the famous maxim of Plato, in that context, namely, in the
sense that;

\bigskip \noindent (8.4)\hfill
\begin{minipage}{11cm}
physical/geometrical is the {\it ``relational''} not the {\it
``spatial''}! In this concern, we thus conclude that {\it
``space''} is the result of {\it laws} (:\;relations/equations),
not the other way around, the latter being thus space-independent
(invariant), which, of course, still delineates (8.3).
\end{minipage}

\bigskip \noindent
Therefore, technically speaking now, the above accentuation of
``$\ca$'' points out, time and again, the {\it significance of}
the {\it ``sheaf of coefficients''}, or even {\it ``structure
sheaf''}, {\it for the ``geometry''} in general! Indeed, the
latter is thus here a {\it spin-off}, in effect, {\it of} $\ca$
(:\;structure sheaf), the quality/effectiveness of our {\it
``geometry''} being, in that sense, {\it directly dependent on
that} particular {\it sheaf} $\ca$. [Thus, take, for instance, the
so-called {\it ``Euclidean spaces''}, globally or not, being
virtually outcomes, speaking analytically (Descartes), of our
classical {\it number fields} (here, {\it constant structural
sheaves}!)].

Consequently, the {\it ``geometry''}, we actually employ, is {\it
``cartesian''}, in spirit, that is, {\it ``analytical''} (consider
``$\ca$'' for instance), not {\it ``Euclidean''}, that might be
viewed, as the {\it ``physical/relational''} one (see (8.3),
(8.4), as above). Thus, we are, in effect, in accord with the very
etymology of the Greek word
``$\gamma\epsilon\omega\mu\epsilon\tau\rho\acute{\iota}\alpha$''
(:\;earth measuring), which, of course, is analytical in the
previous sense; yet, the result of {\it measuring} is a {\it
number} (!) (that entails {\it physics}, according to Feynman). So
it is simply here, where an intrinsic conceptual
perplexity/entanglement is actually entailed, when employing the
notion of {\it ``geometry''}, in the aforementioned point of view;
that is, one is thus confronted with the correspondence ({\it
technical}, in point of fact, {\it identification}),
\[
    euclidean \ \ \longleftrightarrow \ \ cartesian
    \leqno{(8.5)}
\]
As a result, one thus arrives at the {\it Einsteinian}, so to say,
{\it point of view} (cf. [6: p. 164, B.]), in that:

\bigskip \noindent
(8.6)\hfill
\begin{minipage}{11cm}
{\it laws/equations should be universal}, not depending on the
particular type of the {\it ``space''} considered,
\end{minipage}

\bigskip \noindent
the latter being, in view of the preceding, virtually due, for
that matter, entirely, to {\it our} own {\it description} (viz.
the {\it ``cartesian''}, as above, aspect) {\it of the space}!

Accordingly, we are tempting to say, once more (see thus also
(1.1) in the foregoing), that;

\bigskip \noindent
(8.7)\hfill
\begin{minipage}{11cm}
(any!) {\it ``space''}, at all, is delimited, by the {\it
particles} (:\;{\it quanta}) {\it themselves}, to which {\it the
laws} of Physis {\it actually refer}.
\end{minipage}

\bigskip \noindent
Therefore, it would of course be of paramount importance, if any
time we could afford a {\it mechanism} (yet, {\it ``ars
combinatoria''}, \`{a} la Leibniz) that could {\it deal with them}
(the particles) {\it directly}, without thus the intervention of
any supporting/surrounding ``space'' (as it is, for instance, the
case, instead, in the classical theory: CDG).

Now, based on the preceding, we are still led to say that;

\bigskip \noindent
(8.8)\hfill
\begin{minipage}{11cm}
a {\it similar situation} as above {\it holds also true}, which
pertains to the so-called {\it ``quantum deep''}, the {\it
``discrepancies''} revealed, when dealing with it, {\it being
due}, of course, {\it not to the} ({\it ``Euclidean''} (cf. (8.5))
aspect of the) {\it ``space''}, this being actually {\it the same
overall} (!) (Physis is united), but simply to the {\it
``cartesian''} (:\;{\it analytical}, see e.g. $\ca$, yet out {\it
laboratory}) {\it manner} of looking at it.
\end{minipage}

\bigskip \noindent
Consequently, {\it we are} thus {\it trapped} here, time and
again, {\it by the} above (fictitious) {\it ``identification''},
still {\it ours}(!), for that matter, {\it as} it is {\it
indicated} in the foregoing {\it by} (8.5).

On the other hand, classically speaking, {\it the general theory
of relativity is} ({\it ``can be conceived only as''},
A.\;Einstein, loc. cit.) {\it a field theory}, in particular, a
{\it ``continuum theory''} (ibid., pp. 140, 164). Here, {\it
``continuum theory''} is conceived, as that one, that rests on the
{\it ``continuum''}, and this is still in the standard parlance,
the so-called {\it ``space-time} [smooth (!)] {\it continuum''}.
Thus, Einstein himself was looking at the {\it ``quanta''} as {\it
``singularities''} of (viz. something {\it inconsistent with}) the
previous theory, in the sense, in effect, that the relevant
(quantum) theory did not function, within the initial ({\it
``continuous''}) framework.

Thus, here too, one is actually confronted with the aforesaid {\it
entanglement of the perspectives of the notion of ``space''}, as
in (8.5), concerning, at least, the employed herewith {\it
mechanism of ``differential geometry''}; that is, in point of
fact, and {\it this is} here {\it of importance}\,(!),

\bigskip \noindent
(8.9)\hfill
\begin{minipage}{11cm}
one should, instead, refer actually, by the above last term (viz.
{\it ``differential geometry''}), to the intrinsic {\it ``ars
combinatoria''}, \`{a} la Leibniz, this being virtually {\it
independent of any} (smooth)  \linebreak
\end{minipage}

\noindent $ $\hfill
\begin{minipage}{11cm}
{\it ``continuum''} (main moral of ADG, in that context, see also
concluding remarks of the present Section). That is, in other
words, one should refrain here from employing the, so to say, {\it
``Leibnizian mechanism''}, {\it within the wrong scaffolding},
viz. the {\it ``continuum''}(!), as it concerns, in particular,
the {\it quantum theory}.
\end{minipage}

\bigskip \noindent
Yet, to repeat it, due to its particular significance, what
amounts to the same thing, one is thus led to conclude that;

\bigskip \noindent
(8.10)\hfill
\begin{minipage}{11cm}
    the so-called (smooth) {\it``continuum''} is, in effect, {\it
    the wrong scaffolding, as it concerns}, at least, the {\it quantum
    theory}, in order to employ what we may call {\it ``Leibnizian
    mechanism''} (of differential geometry), the latter {\it
    being, anyhow, independent of the former}!
\end{minipage}

\bigskip \noindent
In conclusion, one thus realizes that:

\bigskip \noindent
(8.11)\hfill
\begin{minipage}{11cm}
    even though {\it Nature is united, we employ} for its
    description, depending on the particular case at issue, {\it
    different techniques}, since we are actually trapped by the
    success of our method, referring to the {\it classical
    differential geometry} (CDG), in the case of the {\it
    ``continuum''}, trying thus to alter it, when confronting with
    the {\it ``discrete''} ({\it quantum r\'{e}gime}), while still
    retaining the {\it ``Euclidean''}, in the sense of the {\it
    ``continuum''} aspect of the {\it ``space''}. Hence, the
    apparent differences (:\;{\it ``singularities''} of the
    latter), are actually due to the fact that,

\medskip
\noindent (8.11.1)\hfill
\begin{minipage}{9cm}
    the {\it ``cartesian'' perspective of the ``space'' is}, in
    effect, {\it different in nature, from} that one of the {\it
    ``Euclidean''} point of view, {\it in the sense of} (8.5),
    though both aspects can be, technically speaking, identified
    (\`{a} la Descartes), viz. to the extent that CDG is
    applicable!
\end{minipage}

\medskip \noindent
Notwithstanding,

\medskip
\noindent (8.11.2)\hfill
\begin{minipage}{9cm}
    the {\it innate mechanism} (viz. the aforementioned {\it
    ``Leibnizian''} one) of our {\it method} (:\;differential
    geometry) {\it permeates both aspects}, as they are given by
    (8.5).
\end{minipage}
\end{minipage}

\bigskip \noindent
The content of (8.11.2), as above, is an outcome of ADG. Indeed,
the same follows from several relevant remarks, throughout the
preceding of the present discussion. On the other hand, this point
of view has been recently further corroborated, by two particular
instances of the manner that one can apply ADG, in that context,
which also might, very likely, explain the {\it apparent conflict}
between, generally speaking, the {\it ``continuum''} and the {\it
``discrete''}, pertaining thus to the {\it way of applying}
thereon the {\it standard differential geometry} (of smooth
manifolds); thus, in point of fact, {\it the power} of the latter
{\it is} actually {\it the upshot} of it, viz. the {\it intrinsic}
({\it ``Leibnizian''}) {\it mechanism, not the} particular {\it
source} of it, that is, for the case in focus, {\it the smooth
manifold}. So exactly here appeared lately the aforesaid
vindication/application of the previous claim, by two particular
examples, as mentioned above, viz. that of {\it ``Rosinger's
algebra sheaves''} [31, 24, 25], on the one hand, along with those
of {\it ``Raptis' reticular} (incidence) {\it algebras''} [30, 22,
23], on the other.
\\

{\bf 9.} {\bf Field quantization, in terms of ADG}.$-$ According
to the preceding, the {\it ``space''} on which the ({\it
physical}) {\it objects} (:\;{\it elementary particles}) {\it live
is united} (cf. (8.11), along with (8.3), (8,4)); notwithstanding
that its presence, as a notion, {\it is}, technically speaking,
{\it quite secondary}, given that on the one hand,

\bigskip \noindent
(9.1)\hfill
\begin{minipage}{11cm}
    {\it the} (physical) {\it objects themselves are the
    ``variables'' of the equations of} (the physical {\it laws}
    represented by) {\it the theory} (cf. (8.6), (8.7)), and yet, when
    referring to {\it an elementary particle} as above, we
    actually mean, by virtue of the very definitions, {\it the
    ``quantum''} itself {\it of the } relevant {\it ``field''}
    that is associated with the particle at issue (see also (8.1),
    as well as, A.\;Mallios [20; (1.5)] and [21]),
\end{minipage}

\bigskip \noindent
while on the other, that {\it the} ({\it ``Leibnizian''}) {\it
mechanism} employed (viz. ADG {\it itself}) is {\it quite
independent of any preexistent ``smooth'' structure on the
``space''} in focus.

Within the same vein of ideas, we can still remark here that, as
already mentioned above,

\bigskip \noindent
(9.2)\hfill
\begin{minipage}{11cm}
    our {\it equations} representing the {\it laws of
    Nature}, even in the classical theory (CDG), {\it refer to
    the} (physical) {\it objects themselves} and are virtually
    {\it independent of any} supporting {\it space}
    (see,\linebreak
\end{minipage}

\noindent $ $\hfill
\begin{minipage}{11cm}
    for
    instance, the equations of nowadays {\it ``coordinate free''}
    (classical) differential geometry).
\end{minipage}

\bigskip \noindent
Accordingly,

\bigskip \noindent
(9.3)\hfill
\begin{minipage}{11cm}
    {\it even in the classical theory}, the intervention of {\it
    ``space''} in our equations {\it is absent}.
\end{minipage}

\bigskip \noindent
Of course, this is certainly extremely important, the same
equations (laws) being, {\it otherwise deprived of} their {\it
``geometrical''} (:\;{\it physical/natural}) {\it substance}\,(!).
Indeed, the only participation here of the (cartesian) ``space''
(:\;smooth manifold of CDG) is to supplying us with the {\it
Calculus}, in point of fact, the {\it ``raison d' \^{e}tre''}, and
{\it quintessence},  as well, {\it of} the same {\it Differential
Geometry}, concerning the classical theory. We thus have here,
simply,

\bigskip \noindent
(9.4)\hfill
\begin{minipage}{11cm}
    another {\it justification, from} CDG {\it itself}, that {\it
    ``space'' is} actually {\it irrelevant to the real substance
    of our equations}, the latter {\it referring directly to the
    laws} that govern ({\it relations between}) the {\it physical
    objects} themselves.
\end{minipage}

\bigskip
Now, by looking again at the framework of ADG, thus, in view of
our description therein of {\it elementary particles} (fields), as
pairs
\[
    (\ce, D)
    \leqno{(9.5)}
\]
(see, for instance, [20: (1.8), (1.9)]), in terms of {\it sheaf
theory}, and based further on the respective, within the same
set-up, treatment of {\it ``second quantization''} (see [20],
[21]), one can still remark that,

\bigskip \noindent
(9.6)\hfill
\begin{minipage}{11cm}
    our {\it equations} within the {\it setting of} ADG, are, in
    effect, {\it relativistic}, in character (viz. {\it ``gauge
    independent''}); thus, in point of fact, they are {\it
    ``space-invariant''}, or even, in ADG-parlance, {\it
    ``$\ca$-invariant''}.
\end{minipage}

\bigskip \noindent
On the other hand,

\bigskip \noindent
(9.7)\hfill
\begin{minipage}{11cm}
    based further on (the {\it sheaf-theoretic}, cf., for
    instance, [20], {\it interpretation} of) {\it second
    quantization},

\medskip \noindent
(9.7.1)\hfill
\begin{minipage}{9cm}
    {\it the same equations}, as above, {\it might be construed as being}
    already {\it quantized ones}.
\end{minipage}
\end{minipage}

\bigskip \noindent
We also noted above that {\it our equations are} $\ca$-{\it
invariant}, hence, {\it relativistic} (see (9.6)), alias, by
employing herewith a classical terminology, {\it ``gauge
invariant/in\-de\-pen\-dent''}; yet, in this concern, we should
still remark that;

\bigskip \noindent
(9.8)\hfill
\begin{minipage}{11cm}
    the above {\it ``invariance'' does not} necessarily {\it refer
    to any} surrounding {\it smooth} (e.g. space-time) {\it
    manifold}, as it is usually the case in the classical theory
    (e.g. general relativity, in terms of CDG), {\it but to
    something}, according to the preceding (i.e., to the point of
    view of ADG), {\it more innate}, viz. to our own {\it
    ``arithmetics''} $\ca$, the latter being for that matter
    responsible for the whole edifice, as this is presented, thus
    far. So, in effect, to something {\it more natural}, pertaining
    thus directly to the {\it physical law/relations} of the
    objects at issue, {\it the only parameter} here {\it being}
    ``$\ca$'', {\it with respect} exactly to {\it which the equations
    are invariant}. To repeat it, once more;

\medskip \noindent (9.8.1)\hfill
\begin{minipage}{9cm}
    our {\it equations} are {\it invariant, with respect to} our
    own {\it ``arithmetics''} $\ca$, viz., so to say, relative to
    the same way of our {\it ``measurement''}, this being still
    {\it independent of any space} (Riemann, cf. (1.3)), {\it
    referring} instead {\it directly to the} (physical) {\it
    objects} (e.g. particles) {\it themselves}.
\end{minipage}

\medskip \noindent
    Indeed, all the {\it equations} considered hereto are always
    expressed, {\it exclusively, in terms of sheaf morphisms} between
    the objects (sheaves) involved, the {\it physical objects}
    (particles) at issue  being {\it identified} (e.g. according to
    second quantization) {\it with} appropriate {\it sheaves}.
\end{minipage}

\bigskip \noindent
On the other hand, paraphrasing I.\;Raptis [32], we can ask,

\bigskip \noindent (9.9)\hfill
\begin{minipage}{11cm}
    {\it ``... whether quantizing ... is physically meaningful at all
    ..''},
\end{minipage}

\bigskip \noindent
(our emphasis) which might also be related with C.v.\;Westenholz's
aspect that,

\bigskip \noindent
(9.10)\hfill
\begin{minipage}{11cm}
    {\it `` ... Quantization is provided by the Physical law
    itself''}
\end{minipage}

\bigskip \noindent
(emphasis is ours above; see [39: p. 323]). As a result, and in
connection with our claim in (9.7.1), we can also remark that;

\bigskip \noindent
(9.11)\hfill
\begin{minipage}{11cm}
    {\it the} aforementioned {\it equations}, as in (9.6) and
    (9.7), {\it might} further {\it be viewed}, {\it as} already {\it
    quantized ones, to the extent that they prove to be}, this
    depending thus on the type of the theory employed to their
    deduction, {\it more akin to Physis}, a natural criterion
    thereof, being, of course, dependent on several types of
    occasional inconsistencies appeared, e.g. {\it
    ``singularities''} (always, a {\it spin-off of our theory},
    used in that context).
\end{minipage}

\bigskip \noindent
Now a given {\it equation} may be viewed, as

\bigskip \noindent
(9.12)\hfill
\begin{minipage}{11cm}
    {\it ``more akin to Physis''}, viz. {\it more natural},
\end{minipage}

\bigskip \noindent
whenever it {\it concerns the particle itself} (in vacuo), viz.,
in terms of ADG, the pair,
\[
    (\ce ,D),
    \leqno{(9.13)}
\]
without the intervening of any {\it ``space''} (other, of course,
than the particle at issue), the same being {\it the only
variable} in the equation, that should also be $\ca$-{\it
invariant} (hence, the appearance herewith, as already explained
in the preceding, in place of $D$, actually of $R(D)$).

Thus, an equation, as above, may be construed, in effect, as the

\bigskip \noindent
(9.14)\hfill
\begin{minipage}{11cm}
    {\it quantum field theory} (alias, {\it quantum relativistic})
    - equation, {\it of the particle/quantum}, under consideration.
\end{minipage}

\bigskip \noindent
Furthermore, within the same vein of ideas, we still remark that;

\bigskip \noindent (9.15)\hfill
\begin{minipage}{11cm}
    by considering the aforesaid equations pertaining to a given
    {\it Yang-Mills field} $(\ce, D)$) as before, these are
    actually concerned with what may be construed as the even
    {\it ``formally quantized''} version of the previous field,
    viz. with the following {\it Yang-Mills field},
    \[
        (\ce nd \ce, D_{\ce nd \ce}),
        \leqno{(9.15.1)}
    \]
    which can further be viewed, as a {\it ``matrix
    representation''} of the given one (see e.g [21]).
\end{minipage}

\bigskip \noindent
Consequently, the resulting correspondence, as above,
\[
    (\ce, D) \rightsquigarrow (\ce nd \ce, D_{\ce nd \ce})
    \leqno{(9.16)}
\]
might be considered, as the result of the final action of a {\it
``second-quantization functor''}, acting within the {\it
``Yang-Mills category''} (see also [21], concerning the
terminology applied herewith). In this concern, we also note that;

\bigskip \noindent (9.17)\hfill
\begin{minipage}{11cm}
    {\it the above transition} (9.16) is still encoded in the {\it
    moduli space of} $\ce$,
    \[
        M(\ce):=Conn_A (\ce)/\ca ut \ce =\cc onn_A (\ce)/(\ce nd
        \ce)^{^{\mbox{\bf .}}},
        \leqno{(9.17.1)}
    \]
    the {\it inner structure} of which is further investigated, in
    terms of the same {\it ``matrix representation''} of $\ce$, that is,
    by means of the pair (9.15.1).
\end{minipage}

\bigskip \noindent
Of course this holds true, in particular for the {\it Einstein
equation in vacuo} [19], that is, concerning the relation,
\[
    \crr ic (\ce)=0,
    \leqno{(9.18)}
\]
for a given {\it Yang-Mills field},
\[
    (\ce ,D)
    \leqno{(9.19)}
\]
on $X$, the latter being an ({\it abstract}) {\it Einstein space}.
See loc. cit., or even A.\;Mallios [21], for full details of the
terminology applied herewith.

On the other hand, the whole set-up is, in point of fact, {\it
extremely ``non-linear''} and even {\it ``relativistic''}: Yet,
the {\it ``second quantization functor''},
\[
    (\ce ,D)\rightsquigarrow (\ce nd \ce ,D_{\ce nd \ce})
    \leqno{(9.20)}
\]
supplies a, so to say, {\it ``canonical matrix representation''}
of the initial object, viz. of the ({\it free}) {\it elementary
particle}
\[
    (\ce ,D).
    \leqno{(9.21)}
\]
\indent Now, by looking at (9.18), we see that the same relation
is, in effect, the {\it Yang-Mills equation for the curvature
tensor}. On the other hand, the same meaning of this equation is
actually {\it Machado's} (poetic) {\it verse}:

\bigskip \noindent
(9.22) \hfill
\begin{minipage}{11cm}
    ... paths are made by walking'', viz. the {\it ``traveller}
    [himself] {\it makes the path''}, {\it who}, in point of fact,
    {\it does not feel anything on his trace} ({\it Spur}), viz. he
    feels {\it curvature zero}, during his travel (in vacuo (!), of
    course).
\end{minipage}

\bigskip \noindent
Therefore, {\it it would be, at least strange to think that this
fundamental law of Physis is valid macroscopically only}, and {\it
not} in {\it the ``quantum deep''}, {\it as well} (!).
Accordingly, we are led to claim that:

\bigskip \noindent
(9.23)\hfill
\begin{minipage}{11cm}
    {\it (9.18) is Einstein's quantum-equation} too,
    viz. e\;o\;\;i\;p\;s\;o the {\it quantized Einstein's equation}
    ({\it in vacuum}).
\end{minipage}

\bigskip \noindent
Finally, the relation,
\[                      %(9.24)
    \crr ic:=tr \circ R(., s)t
    \leqno{(9.24)}
\]
is an $\ca$-{\it morphism of $\Om^2 \equiv \wedge^2 \ce$ in}
$\ca$, so that, according to (9.18),

\bigskip \noindent (9.25)\hfill
\begin{minipage}{11cm}
    Einstein's equation (in vacuum) is just the {\it kernel} of
    the same operator (:\,$\ca$-{\it morphism}, as before), that
    also delimits the corresponding {\it Einstein} (:\,{\it
    solution}) {\it space} of the equation at issue (see [21]).
\end{minipage}

\bigskip \noindent
On the other hand, due to the {\it gauge invariance of the trace}
operator, one further concludes that:

\bigskip \noindent (9.26)\hfill
\begin{minipage}{11cm}
    {\it Einstein's equation} (in vacuo) {\it retains its meaning}
    still {\it on the moduli space of a} given {\it solution} of
    it.
\end{minipage}

\bigskip \noindent
Yet, concluding the present account, we should like to underscore,
once more, a {\it mathematical-physical correspondence}, that
actually permeated our whole discussion herewith, mainly motivated
by a relevant terminology originally employed by Yu.I.\;Manin
[26], and further justified by what we may call {\it ``Selesnick's
correspondence principle''} [21: Chapt. II], in effect, a
formulation of {\it second quantization}; so based on the
aforesaid correspondence, we can still say that, in the same way
that,

\bigskip \noindent (9.27)\hfill
\begin{minipage}{11cm}

    \medskip \noindent (9.27.1)\hfill
    \begin{minipage}{9cm}
        {\it a sheaf is its sections}
    \end{minipage}

    \medskip \noindent
    see [18: Chapt. I; Section 3], thus, also;

    \medskip \noindent (9.27.2)\hfill
    \begin{minipage}{9cm}
        {\it an elementary particle is its states},
    \end{minipage}

    \medskip \noindent
    viz. {\it the sections} (in view of (9.27.1), as above) {\it
    of the sheaf}, that is {\it associated with it}, according to
    the aforementioned correspondence.
\end{minipage}

\bigskip \noindent
Thus, still finally, we are led, time and again, to emphasize
that;

\bigskip \noindent (9.28)\hfill
\begin{minipage}{11cm}
    {\it locally} our {\it descriptions} (of the physical events)
    {\it are made through sections} of our (generalized) {\it
    ``arithmetics''}, or of what we have called in the preceding
    {\it ``structure sheaf''} $\ca$. So it is, via this same
    structure sheaf $\ca$, that one should also {\it describe} the
    {\it ``space''}, as we actually did, strongly motivated, in
    effect, thereto, by ADG itself, by identifying (cf.\;(1.1)) the
    former with the (physical) objects $\longleftrightarrow$ (vector)
    sheaves (cf. (9.27)), that live on it.
\end{minipage}

\bigskip \noindent
In this concern, we should still note that;

\bigskip \noindent (9.29)\hfill
\begin{minipage}{11cm}
    even in the {\it classical theory} the description of the
    ({\it ``physical''}) {\it space} is actually made (\`{a} la
    Descartes), via the corresponding therein {\it structure
    sheaf}
    \[
        \ca \equiv \cc^{\infty}_{X}.
        \leqno{(9.29.1)}
    \]
\end{minipage}

\bigskip \noindent
However, the extremely nice properties of the latter sheaf in the
classical case are actually due to the

\bigskip \noindent (9.30)\hfill
\begin{minipage}{11cm}
    {\it algebraic-topological} (hence, even {\it differential
    analytic}\,-\,Hilbert) consequences of the {\it basic
    arithmetics}, viz of our standard {\it numerical fields}, we
    employ therein, hence, the analogous properties of the {\it
    basic} ($\br$-, or $\bc$-) {\it algebra} (of global sections)
    \[
        \ba \equiv \ca (X) \equiv \cc^{\infty}_{X}(X).
        \leqno{(9.30.1)}
    \]
\end{minipage}

\bigskip \noindent
As a result, one thus comes to the tempting conclusion that, it
seems as that,

\bigskip \noindent (9.31)\hfill
\begin{minipage}{11cm}
    {\it it is more natural to think} (:\;make arguments, ideas),
    in terms of {\it algebra, rather than geometry}.
\end{minipage}

\bigskip \noindent
The above reminds us of a relevant statement of C.v.\;Westenholz
[39: p. 180], in the sense that;

\bigskip \noindent (9.32)\hfill
\begin{minipage}{11cm}
    ``... most physical observables must adequately be described in
    terms of exterior forms and not by vectors''.
\end{minipage}

\bigskip \noindent
Yet, even of Einstein's [6: p. 166],

\bigskip \noindent (9.33)\hfill
\begin{minipage}{11cm}
    ``... find a purely algebraic theory for the description of
    reality''.
\end{minipage}

\bigskip \noindent
The reward or the advantage from such a choice, as in (9.31),
could be, at least, that, by in refernce for instance to ADG;

\bigskip \noindent (9.34)\hfill
\begin{minipage}{11cm}
    the algebra (-way of thinking) might contain (viz. engulf)
    {\it geometrical discrepancies}, alias {\it
    ``singularities''}, by remaining, notwithstanding, {\it
    still operative---that is, still work}\,(!) In other words, {\it algebra proves
    thus to be more flexible},
    being, therefore, able to permeate, go through, or even
    circumvent, difficulties (:\;{\it ``singularities''}), {\it
    that} the standard point of view of {\it geometry simply cannot} \,!
\end{minipage}

\bigskip \noindent
So we can think here of the famous {\it ``Plank scale''}, that, of
course, it is {\it not, a matter of analysis/algebra} (think of
Dirac's aphorism, see thus (3.9)), but rather of the particular
manner (viz. still classical differential geometry), that it is
here undertaken to exploit {\it ``geometry''}.

All in all, we are very tempted to say that;

\bigskip \noindent (9.35)\hfill
\begin{minipage}{11cm}
    any time we have an extremely efficacious {\it ``geometrical
    mechanism''} (see e.g. classical differential geometry), it
    is very likely that there will be hidden therein (or even,
    it would at least be worthwhile to look for) an {\it
    inherent} ({\it deeper}) {\it algebraic one} that should
    actually be the {\it ``catalyst'' to the former}, making thus
    the things to work {\it without necessarily} the presence of the initial {\it
    ``geometrical scaffolding/panoply''}. (Compare, for instance,
    the numerously mentioned in the preceding Leibniz's {\it
    ``ars combinatoria''}, or even the same framework of ADG).
\end{minipage}

\bigskip \noindent
Nevertheless, one should then further look for the appropriate,
very likely lurking therein, {\it ``representation theorem''},
alias, {\it ``Gel'fand duality''}\,(!), a fact, that it would then
actually entail, when {\it properly envisaged}, a {\it converse to
the preceding statement}, as in (9.35)\,(!) (see, for instance,
what happened in nowadays Algebraic Geometry).

Thus, one is virtually led, at the very end, back again to Sophie
Germain, in that;

\bigskip \noindent (9.36)\hfill
\begin{minipage}{11cm}
    {\it ``L' alg\`{e}bre n' est qu' une g\'{e}om\'{e}trie \'{e}crite, la
    g\'{e}om\'{e}trie n' est qu' une alg\`{e}bre figur\'{e}e.''}
\end{minipage}

\bigskip \noindent
(emphasis above is ours; cf. S.\;Germain in {\it
``Pens\'{e}es''}).

Accordingly, based finally on (9.35)/(9.36), we may wind up, by
still saying that:

\bigskip \noindent
(9.37)\hfill
\begin{minipage}{11cm}
    an established {\it ``geometrical theory'' entertaining},
    however, {\it ``singularities''}, might be effectively viewed just
    as an {\it insufficiently rendered}, an ill-represented in that respect,
     {\it ``algebraic mechanism''}.
\end{minipage}

%%%%%%%%%%%%%%%%%%%%%%%%%%%%%%%%%%%%%%%%%%%%%%%%%%%%%%%%%%%%%%%%%%%%%%%%%%%%
\begin{center}

\end{center}

\end{document}